\documentclass[nofootinbib,notitlepage,longbibliography]{revtex4-1}
\usepackage{wasysym}
\usepackage{latexsym}
\usepackage{graphicx}
\usepackage{dcolumn}
\usepackage{bm}
\usepackage{booktabs}
\newcommand{\tabitem}{~~\llap{\textbullet}~~}

\begin{document}

\title{Nuclear spin correlations and collective excitations in supercritical H$_2$.}

\author{Raina J. Olsen$^1$}
\email{olsenrj@ornl.gov}
\author{Jon W. Taylor$^2$}
\author{Cristian I. Contescu$^1$}
\author{James R. Morris$^1$}
 	
\affiliation{$^1$Materials Science and Technology Division, Oak Ridge National Laboratory, Oak Ridge, TN 37831, USA} 
\affiliation{$^2$ISIS Spallation Neutron Source, STFC, Rutherford Appleton Laboratory, Didcot OX11 0QX, United Kingdom}

\maketitle

\textbf{A longstanding challenge in quantum computing is to find qubits that interact strongly with one another, but weakly with their environment to prevent decoherence, properties difficult to find in a single physical implementation.\cite{ladd2010quantum} Present technologies use strongly interacting qubits for two-qubit gates, \cite{veldhorst2014two,barends2014} while weakly interacting nuclear spins are useful for one-qubit gates \cite{pla2013high} and coherent memory.\cite{morton2008solid,saeedi2013room} Nuclear spins are known to experience spontaneous long-range correlations only below 2.5 mili-Kelvin in superfluid $^3$He \cite{leggett1975theoretical}. Here we present the first evidence of nuclear spin coupling in molecular hydrogen (H$_2$) at 74-92 Kelvin using neutron scattering, showing a fundamental change in nature from the incoherent scattering universally expected from hydrogen, which reflects single particle properties of uncorrelated nuclear spins \cite{YandK,lovesey}, to coherent, with a peak materializing on the elastic line \cite{roinel1978first,oja1997nuclear}  indicating H$_2$-H$_2$ nuclear spin correlations.  In this novel phase, the dynamic response of the system also changes nature, and collective excitations with an effective mass of nine H$_2$ are observed with inelastic scattering at momentum transfers up to 37 \AA$^{-1}$, corresponding to length scales smaller than the H-H bond, where previous experiments have always found single atom excitations \cite{DINSrev, andreani1998deep, greatFit,bafile1998deep,wang1991momentum, langel1988inelastic,bafile1996density, herwig1990density}. This novel behavior has only been observed from H$_2$ within the subnanometer sized graphitic pores of a carbon material \cite{thesis}, marking the first demonstration that a confined materials environment can be used to control nuclear spin correlations. As such, these results show that it may be possible to engineer systems of interacting nuclear spin qubits for error correction and two-qubit gates. }

 \begin{figure}[htb!]
  \centering
\includegraphics[width=0.49\textwidth]{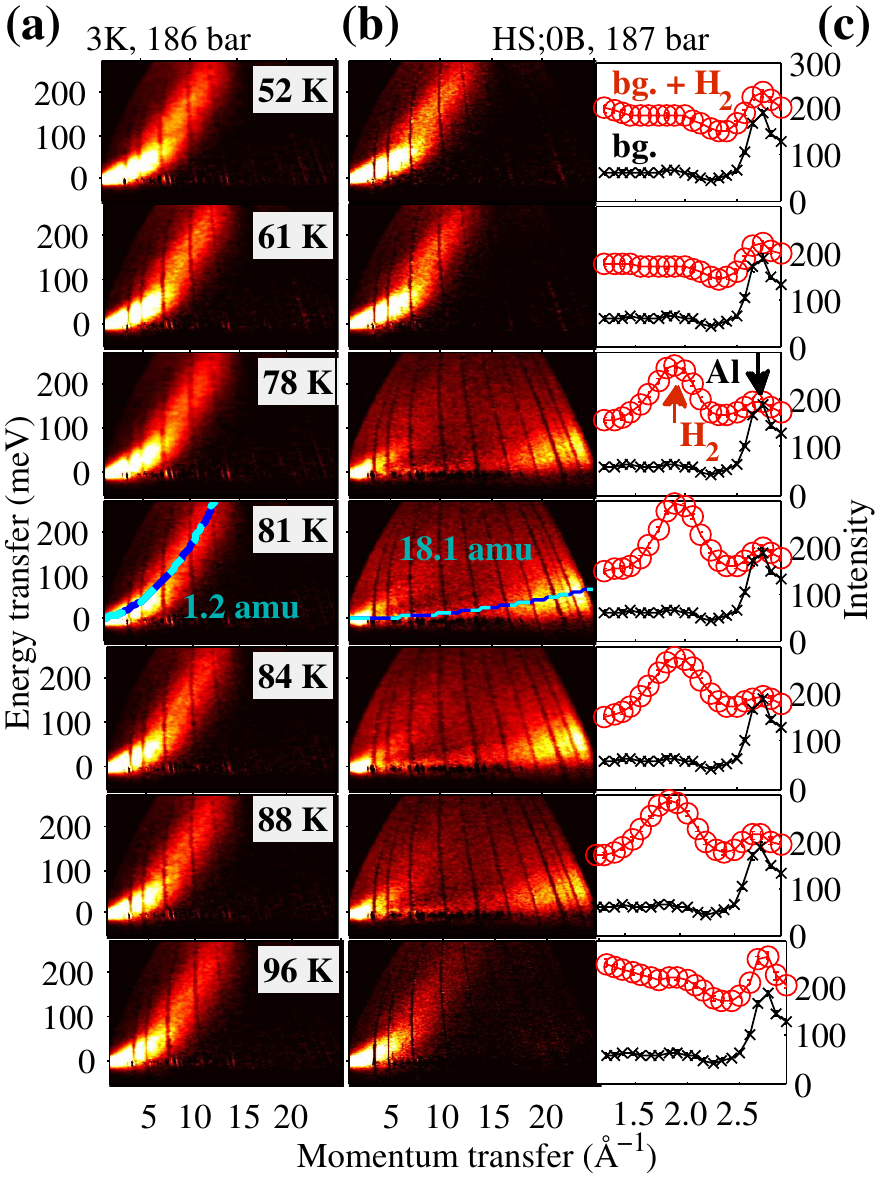}
\caption{Inelastic neutron scattering from adsorbed H$_2$ showing a phase change as a function of sample and temperature. (a) H$_2$ adsorbed in control sample `3K' at a nominal pressure of 186 bar.  The background scattering from the system with no H$_2$ has been subtracted.  (b) H$_2$ adsorbed in sample of interest `HS;0B-3' at a nominal pressure of 187 bar with background subtracted.  All spectra in panels (a) and (b) are shown using the same intensity color scale.    (c)  Data from the elastic line for H$_2$ in `HS;0B-3' summed over $\hbar \omega=\pm 10$ meV, without background subtraction, shown in comparison to the background. }
\label{spec}
\end{figure}

Figure \ref{spec} shows a novel phase change in H$_2$ observed with inelastic neutron scattering (INS). Figs. \ref{spec}(a) and \ref{spec}(b) show INS spectra collected at a series of temperatures between 52--96 Kelvin (K) from high pressure H$_2$ adsorbed in two different nanoporous carbons: `3K', the name of an activated carbon used here as a control, and `HS;0B-3', a locally graphitic carbon with pores having sub-nanometer widths.\cite{thesis}  Materials characterization data on these carbons are shown in the Extended Data (ED) Fig. \ref{mat}.   \footnote{This manuscript has been authored by UT-Battelle, LLC under Contract No. DE-AC05-00OR22725 with the U.S. Department of Energy. The United States Government retains and the publisher, by accepting the article for publication, acknowledges that the United States Government retains a non-exclusive, paid-up, irrevocable, world-wide license to publish or reproduce the published form of this manuscript, or allow others to do so, for United States Government purposes. The Department of Energy will provide public access to these results of federally sponsored research in accordance with the DOE Public Access Plan (http://energy.gov/downloads/doe-public-access-plan). }

At the highest and lowest temperatures, the dynamic response of the H$_2$ in these two carbon materials is nearly identical.  At any given value of the momentum transfer ($Q$) from the neutron to the system under study, the peak energy transfer ($\hbar \omega$) is  equal to the kinetic energy $(\hbar Q)^2/2M$ of a particle with mass $M=1.2$ amu, roughly equal to the mass of a single hydrogen atom $M=1.0$ amu.  This result is consistent with a system of uncorrelated H$_2$ molecules, and fits to these spectra using an H$_2$ molecular model \cite{YandK,greatFit,langel1988inelastic,bafile1998deep} are shown in the ED Fig. \ref{fitf} and described in the Supplemental Information (SI).   Fluid hydrogen has been extensively studied by INS, and this H recoil has always been  observed.\cite{DINSrev, andreani1998deep,greatFit, bafile1998deep,wang1991momentum, langel1988inelastic, bafile1996density}  Even when the H atoms are part of a larger molecule, such as water (H$_2$O),\cite{pietropaolo2008,garbuio2007proton,waterQEdins} or when the H$_2$ forms a solid,\cite{DINSrev,langel1988inelastic,bafile1996density,herwig1990density}  recoil of single H atoms has always been observed.

\begin{figure}[htb!]
  \centering
\includegraphics[width=0.49\textwidth]{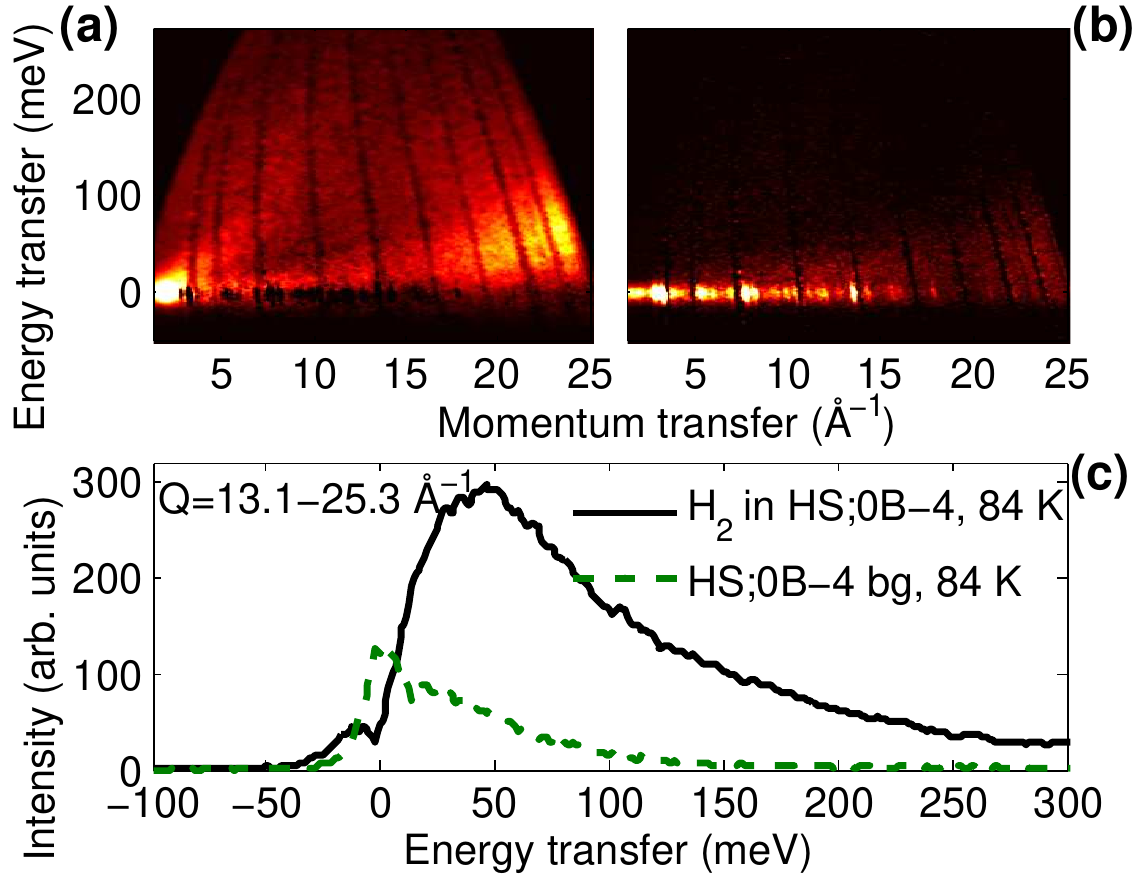}
\caption{Inelastic neutron scattering data (a) from H$_2$ in HS;0B after background subtraction, shown in comparison to (b) the background scattering from the carbon+sample can+quartz wool, showing free recoil of the aluminum ($M=26.98$ amu).  Both spectra are shown with the same intensity color scale.  (c)  Sum over $Q$=13.1--25.3 \AA$^-1$ for the same spectra shown in panels (a) and (b). Not only is the scattering from the novel phase of H$_2$ larger in intensity than the background, but it also peaks at a higher energy, corresponding to a smaller mass than the aluminum. }
\label{bg}
\end{figure}

The scattering from H$_2$ adsorbed in the control sample, `3K', changes only slightly with temperature.  But for H$_2$ in `HS;0B-3', the scattering shows an abrupt change, which occurs solely as a function of changes in temperature, with the novel phase observed between 74--92 K.  Two striking changes in the measured spectra are observed.  Firstly, the effective recoil mass changes from 1.2 amu to 18.1$\pm$0.6 amu.  In addition, a new peak at 1.9 \AA$^{-1}$ appears on the elastic line, shown in Fig \ref{spec}(c).  The novel elastic peak is large compared to the background elastic scattering in the same range of $Q$ collected from the system before the addition of H$_2$, which includes scattering from the carbon and the aluminum pressure cell.  In addition, the maximum intensity of the high mass recoil  is $\sim$5 times larger in intensity than the background scattering in the same region (see ED Fig. \ref{bg} for a direct comparison).    

 \begin{figure}[htb!]
  \centering
\includegraphics[width=0.49\textwidth]{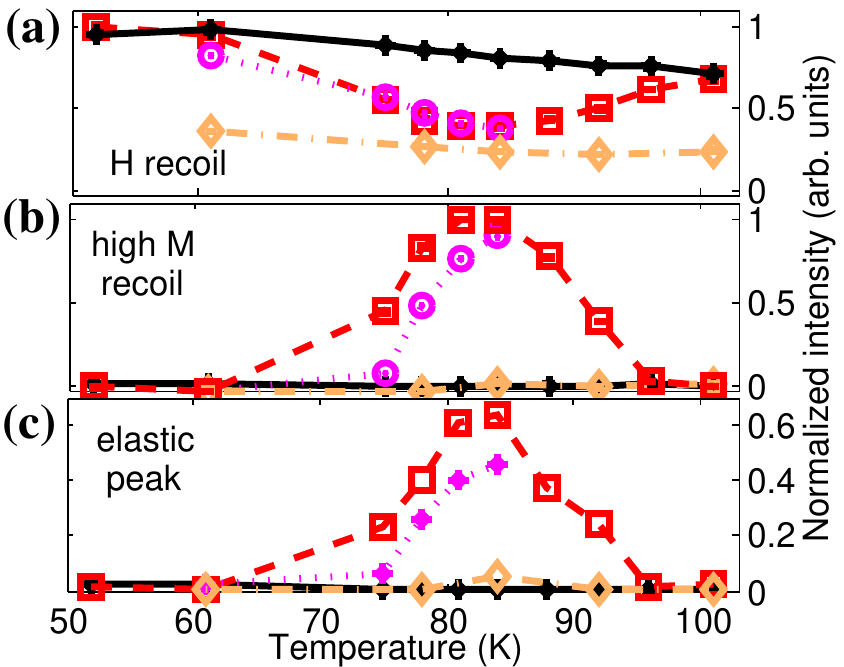}
\caption{Intensity of inelastic neutron scattering spectra within different ranges of $Q$, $\hbar \omega$ from adsorbed H$_2$ as a function of temperature, pressure, and carbon adsorbent. (a) Intensity of 1 amu hydrogen recoil, summed from $\hbar \omega$=14.3--58.8 meV, $Q$=2.5--6.8 \AA$^{-1}$. (b)  Intensity of high mass recoil, from $\hbar \omega$=14.3--73.7 meV, $Q$=18.0--23.5 \AA$^{-1}$. (c) Area under the elastic peak that appears with the high mass recoil, obtained by fitting a linear background plus a Guassian to the elastic scattering.  Data in panels (a) and (b) are normalized to the summed intensity of H recoil at $P$=187 bar and $T$=52 K, and error bars are shown, but are smaller than symbol sizes. Legend: H$_2$ in `HS;0B-3' at P=187 bar: \emph{Red} $\Square$, P=123 bar: \emph{Magenta} $\Circle$, P=30 bar: \emph{Orange} $\Diamond$, and H$_2$ in `3K' at P=187 bar: \emph{Black} $+$. }
\label{intensity}
\end{figure}

To understand the thermodynamic boundaries of the observed phase transition, we show scattering intensities as a function of temperature, pressure, and carbon adsorbent in Figure \ref{intensity}.  The high mass recoil and new elastic peak are always observed together, and only at high H$_2$ pressure in `HS;0B-3'.  Their appearance strictly corresponds to a proportional decrease in the intensity of the $\sim$1 amu H recoil.  In contrast, no high mass recoil or elastic peak is observed at any temperature from H$_2$ in the control sample `3K' or in `HS;0B-3' at low pressure.   Given this observation of several reversible transitions into and out of the novel phase and the large intensity of the novel features relative to the background scattering, we comfortably associate the appearance of high mass recoil and simultaneous decrease in H recoil with a change in phase of the H$_2$ in the system.

Observation of a strong elastic neutron scattering peak from a system of H$_2$ is unprecedented because hydrogen is a strong incoherent scatterer, with a ratio of incoherent to coherent scattering of $\sigma_I/\sigma_C=45.7$, and so produces featureless scattering on the elastic line.  The expectation of featureless scattering can be derived from the assumption that the measured incoherent scattering, $S_I(\vec{Q},\omega)$,  is equal to the time-dependent Fourier transform of the single particle density correlation function,
\begin{equation}
\label{selfcor}
S_I(\vec{Q},t)=\langle e^{i \vec{Q}\cdot \vec{\hat{r}}_j}   e^{-i\vec{Q}\cdot \vec{\hat{r}}_j(t) } \rangle, 
\end{equation}
where $\vec{\hat{r}}_j(t)$ is the position operator of atom $j$ at time $t$ and $\langle \dots \rangle$ represents the thermal average. From Eq. \ref{selfcor}, one can derive the zeroth moment and first moment sum rules,\cite{excitations,lovesey}   
\begin{eqnarray} 
\label{sum0}
\int_{-\infty}^{\infty} S_I(\vec{Q},\omega) d\omega &=&1, \\
\label{sum1}
\int_{-\infty}^{\infty} S_I(\vec{Q},\omega) \omega d\omega &=&\frac{\hbar Q^2}{2 M},
\end{eqnarray}
The zeroth moment sum rule originates from the fact that  Eq. \ref{selfcor} is a density self-correlation function and the particle must always be somewhere, and is an exact prediction of a featureless elastic line.  The first moment sum rule means that the average energy transfer is given by the kinetic energy of a single atom at the given momentum transfer.  Both of these rules are followed  here in the normal phase of H$_2$, as well as in all previous neutron scattering studies of H$_2$ \cite{DINSrev, andreani1998deep, greatFit,bafile1998deep,wang1991momentum, langel1988inelastic,bafile1996density, herwig1990density}.  

But Eq. \ref{selfcor} is an approximation.  The measured scattering is actually the Fourier transform of the nuclear spin pair correlation function, \cite{lovesey} 
\begin{equation}
\label{spincor}
S_I(\vec{Q},t)=\sum_{j,j'}\frac{1}{i(i+1)}  \langle e^{i \vec{Q}\cdot \vec{\hat{r}}_j}  \vec{\hat{I}}_j \cdot \vec{\hat{I}}_{j'} (t)  e^{-i\vec{Q}\cdot \vec{\hat{r}}_{j'}(t) } \rangle, 
\end{equation}
where $ \vec{\hat{I}} $ is the nuclear spin operator.  In most system nuclear spins of different atoms in the system are uncorrelated, in which case $\langle \vec{\hat{I}}_j \vec{\hat{I}}_{j'}(t) \rangle=i(i+1) \delta_{jj'}$ and Eq. \ref{spincor} reduces to Eq. \ref{selfcor}.

It's so rare for systems to have correlated nuclear spins, that $S_I(\vec{Q},\omega)$ is simply called incoherent scattering, rather than nuclear spin dependent scattering.  But the strong elastic peak and high effective mass of the inelastic excitations found in the present work defy both the zeroth and first moment sum rules in Eqs. \ref{sum0}--\ref{sum1}, and thus indicate that the system must have some type of magnetic order with correlated nuclear spins in the novel phase.  Since nuclear spins interact very weakly, the only other systems in which long-range correlated nuclear spins have been observed as they are here with `incoherent' or spin-dependent elastic neutron scattering are those in which nuclear spin ordering is driven by a magnetic field through a hyperfine interaction with the electronic degrees of freedom at mili-Kelvin temperatures.\cite{roinel1978first,oja1997nuclear}

It is well known that short-range nuclear spin correlations occur between the two identical H atoms  in an H$_2$ molecule as a result of exchange symmetry.  \cite{lovesey,YandK}  This results in two distinct forms of hydrogen: para-H$_2$ with a net nuclear spin $S=0$, and ortho-H$_2$ with a net nuclear spin $S=1$.  But the observed elastic peak  is at $Q{\sim}$1.9 \AA$^{-1}$, corresponding to a spacing of 3.3 \AA, a separation at which the H$_2$-H$_2$ potential is strongly attractive,\cite{wang} and much larger than the H--H bond distance of 0.71 \AA.  Thus the results indicate that there are H$_2$-H$_2$ nuclear spin correlations in the novel phase.  While the momentum resolution of this experiment is not ideal, the resolution corrected width of the elastic peak is consistent with a correlation length of $\sim$35 \AA, larger than the correlation length of $\sim$11 \AA \ in bulk H$_2$,\cite{paircor} thus indicating that the spin correlations are not simply short-range.

Normally a phase transition that occurs as the temperature is lowered indicates passage of the system to a more ordered state, and we expect this order to persist as the temperature is lowered further.  If the higher temperature bound of the observed phase at 92 K is associated with magnetic order of the H$_2$ nuclear spins, then why do we see this order dissapear at 74 K?  

While nuclear spins interact weakly, the net H$_2$ spin is coupled to its orbital angular momentum, which has a large energy dependence.   Because the H protons are fermions and the wavefunction of identical fermions must be antisymmetric under exchange of the particles, the aligned proton spins of ortho-H$_2$ must be combined with a wavefunction which is antisymmetric in space, $J=1,3,\dots$, while anti-aligned spins of para-H$_2$ must be combined with a wavefunction which is symmetric in space, $J=0,2,\dots$, where $J$ is the quantum number of the orbital angular momentum describing the orientation of the molecular axis.   The $J=1$ ortho state is 171 K above the $J=0$ para state.  At room temperature, bulk equilibrium H$_2$ is $\sim$75\% ortho, but passes from mostly ortho to mostly para at $\sim$77 K, which is quite close to the lower temperature bound of the observed phase transition at 74 K. Thus we hypothesize that the observed magnetic order occurs only in ortho-H$_2$ with $S=1$, and this magnetic ordering is broken by the increasing concentration of para-H$_2$ with $S=0$ as the temperature is lowered.

At the temperatures of this experiment, ortho-H$_2$ occupies only the lowest $J=1$ rotational state, with the $J=3$ state at 1026 K.  There are a total of 9 accessible states of ortho-H$_2$ with $J=1,S=1$ , corresponding to the combinations of nuclear spin states $S_z=-1,0,1$ and  angular momentum states $J_z=-1,0,1$, where $z$ is an arbitrary projection axis.  Thus with antiferromagnetic ordering in a system of ortho-H$_2$ we might expect formation of a nonet (a group of 9) with a net spin $\mathcal{S}=0$. This interpretation is consistent with the observed recoil mass of 18.1 amu, which is equal to the mass of 9 H$_2$ molecules.

 \begin{figure}[t!]
  \centering
\includegraphics[width=0.49\textwidth]{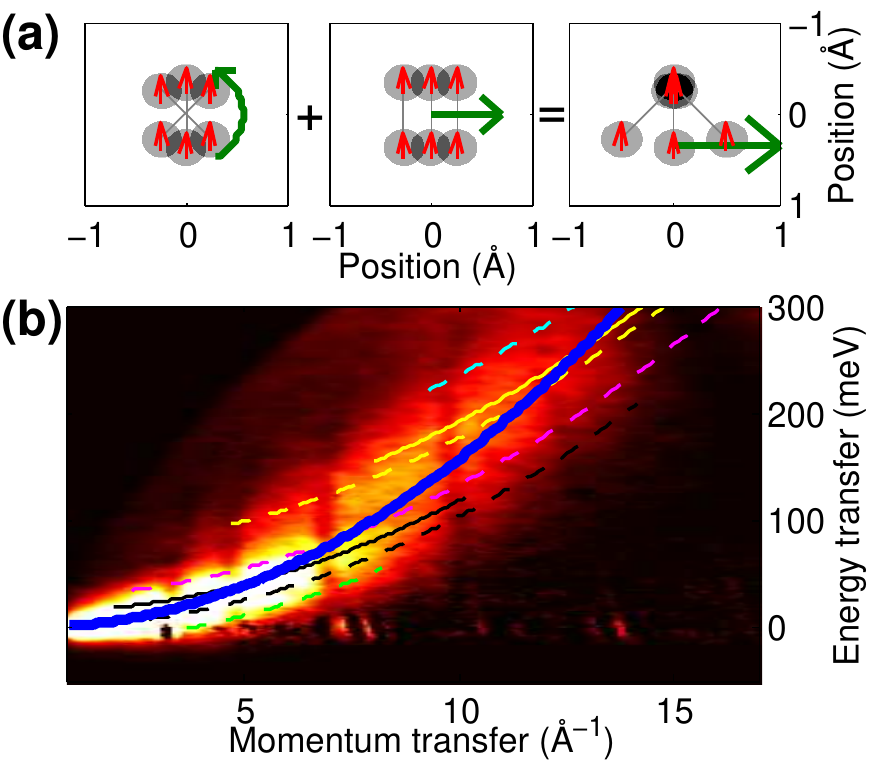}
\caption{Understanding the observation of $\sim$1 amu H recoil in DINS spectra from 2 amu H$_2$.  (a) Cartoon depicting the observed transitions.   A combination of rotational motion of the molecule and translational motion of the center of mass results, on average, in a rolling type motion in which one H atom moves rapidly while the other is nearly fixed.  The spins of the atoms, depicted by red arrows, stay fixed.  (b)  Representative spectrum from adsorbed H$_2$.  A thin recoil line with $M$=2 amu is plotted for each observable rotational transition extending from the energy of the transition. Recoil lines are shown only in the region where the rotational form factor $F_{JJ'}^{H_2}(Q)$ for the given transition is significant.  A thick recoil line with $M$=1.2 amu represents the sum result.   }
\label{H2H}
\end{figure}  

We must emphasize that the observed transition in recoil mass cannot simply be explained by formation of any $\sim$18.1 amu H containing molecule, as the recoil mass derived in Eq. \ref{sum1} is associated with nuclear mass rather than molecular mass.  H recoil with mass $\sim$ 1 amu is observed in fluid\cite{DINSrev, andreani1998deep,greatFit, bafile1998deep,wang1991momentum, langel1988inelastic, bafile1996density}  and solid H$_2$\cite{DINSrev,langel1988inelastic,bafile1996density,herwig1990density}  (molecular mass 2.0 amu) and even in H$_2$O\cite{pietropaolo2008,garbuio2007proton,waterQEdins} (molecular mass 18.0 amu).  Given that the binding energies of these molecules, $\sim$5 eV, are much larger than the neutron incident energy used here, 400 meV, it is perhaps surprising that recoil of the H atom is observed.  This occurs because the neutron tends to excite both rotational and translational excitations of the molecule at the same time.  A given rotational excitation corresponding to a change in rotational velocity $\Delta v_r$ is likely to be observed only at certain values of $Q$ according to the form factor (or scattering probability) $F_{JJ'}^{H_2}(Q)$, at which a given translational motion with a change in the velocity $\Delta v_{cm}$ is also likely to be excited.  On average, $\Delta v_r \sim\Delta v_{cm} $, with the net result a type of rolling motion in which only one atom is excited, as depicted in Figure \ref{H2H}.  Here, we observed $M=$1.2 amu, but $M\rightarrow1.0$ as the neutron incident energy increases.  

\begin{figure}[t!]
  \centering
\includegraphics[width=0.49\textwidth]{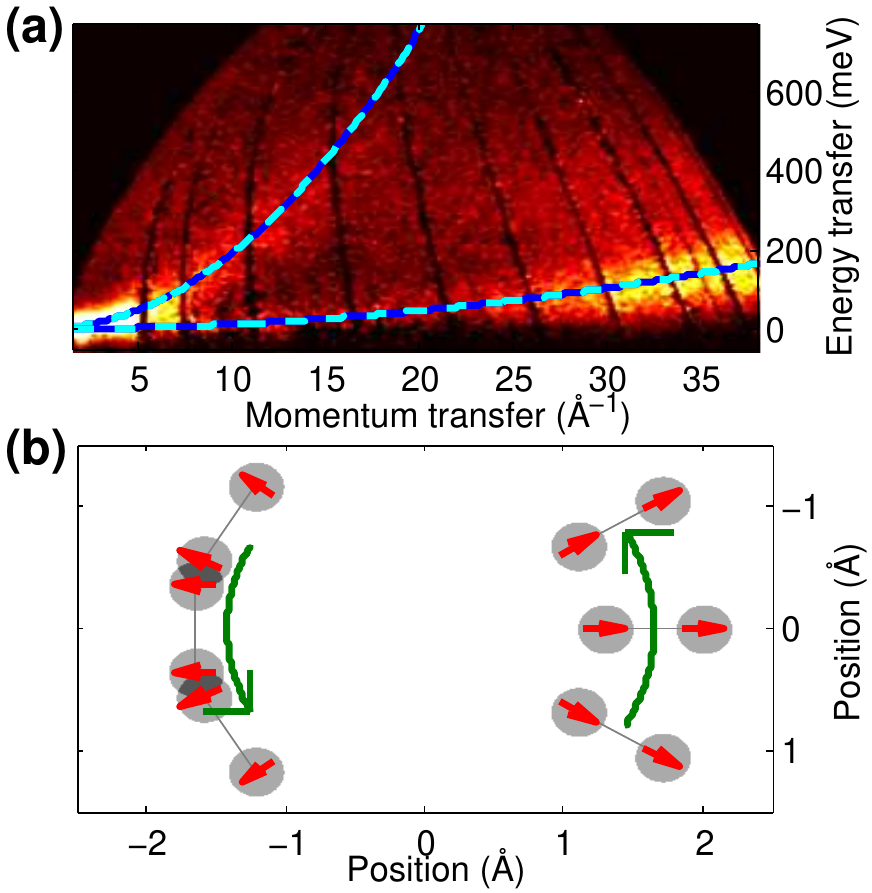}
\caption{Interpretation of the observed 18.1 amu recoil. (a)  Spectrum collected in the novel phase with an incident neutron energy of 1000 meV.  (b) Cartoon depicting two ortho-H$_2$ molecules in a state in which their spins, orientations, and positions are coupled with one another.  The spins of the atoms are depicted by red arrows, and depend on the orientation of the molecular axis.}
\label{H218}
\end{figure}

This picture of a rolling motion in which only one atom is excited is valid only under the assumption that the H$_2$ spin, rotational, and translational degrees of freedom can be treated independently (except with the condition that $J=2n+S$ where $n$ is an integer, as noted above), an assumption which is used to calculate the exact scattering probabilities $F_{JJ'}^{H_2}(Q)$ \cite{YandK}.   The highly novel collective recoil observed with a completely different $F_{JJ'}^{H_2}(Q)$ here means that the spin degrees of freedom must be coupled to the rotational and translational degrees of freedom in some way, as is depicted in Fig. \ref{H218}(b). Until a quantitative theory is developed for this system this is all that can be definitively concluded from the observation of high mass recoil.

But a qualitative analysis may offer insight into possible explanations for the observed behavior.  As shown in Figure \ref{H218}(a), the high mass recoil persists as high as $Q$=37 \AA$^{-1}$.  Because this corresponds to a length scale much smaller than the distance between atoms, the semi-classical impulse approximation normally applies, in which the neutron interacts over such small length and time scales that only one atom is excited over a short distance and the other atoms do not have time to respond to its change in motion.  Observation of collective excitations in this regime seems to be consistent with the presence of long-range correlations in position, in which the motions of atoms in the system that are far apart are directly correlated and respond immediately.  

Long range velocity correlations are a feature of superfluidity, but the temperatures at which we observed this novel phase are an order of magnitude larger than in any known superfluid.  We have previously predicted high temperature superfluidity in a system of  light strongly interacting spinless particles ($^4$He, para-H$_2$)  confined in a periodic potential with a lattice spacing equal to the interparticle spacing at which the intermolecular interaction is strongly attractive, consistent with the 3.3 \AA \ spacing measured here.\cite{HTS}  Pores composed of two parallel layers of graphene with a 0.8--0.9 nm spacing provide the necessary environment, materials properties which are consistent with those of the sample of interest used in the present study.  In superfluid $^3$He,  in which the fermionic atoms forming a Cooper pair have ferromagnetically ordered nuclear spins, collective pair excitations with coupled spin and translational motion have been predicted \cite{wolfle1977collisionless} and observed.\cite{ling1987ultrasonic,davis2008discovery}   Thus the correlated nuclear spins and collective recoil with a mass of nine H$_2$, consistent with antiferromagnetic ordering of the bosonic nuclear spins of ortho- H$_2$, observed here are qualitatively consistent with known phenomena in superfluids.

While we cannot yet claim any identification of the phase of matter exhibited by the system studied here, the starkly unique observation of collective excitations and strong nuclear spin correlations between H$_2$ molecules at temperatures far exceeding those of similar systems sheds light on exciting new avenues of study into methods to manipulate and control quantum information stored in nuclear spins.

\section*{Acknowledgments}
This work was partially supported by the U.S. DOE-EERE Postdoctoral Research Awards under the EERE Fuel Cell Technologies Program.  Experiments at the ISIS Pulsed Neutron and Muon Source were supported by a beamtime allocation from the Science and Technology Facilities Council.  Samples were provided by the ALL-CRAFT group at the University of Missouri.  X-ray research was conducted at the Center for Nanophase Materials Sciences and microscopy research was conducted at the Oak Ridge National Laboratory’s SHaRE User Facility, both of which are DOE-BES User Facilities.  JRM acknowledges support from DOE’s Office of Science, Basic Energy Sciences, Materials Science and Engineering.  The authors thank M. B. Stone and G. Siopsis for discussion.

\clearpage

\section{Methods}

Experiments were conducted using H$_2$ adsorbed in powdered nanoporous carbons: `HS;0B-3,' a carbon synthesized by the pyrolysis of Saran (PVDC)\cite{thesis} in which the high mass recoil is observed, and a material called `3K,' a KOH activated carbon used here as a control.   In general, Saran carbons are classified as non-graphitizable isotropic carbons,\cite{marsh2006activated} meaning that there is, on average, no preferred direction for graphene planes to lie and the carbon do not become graphitic on heating to 2000 $^\circ$C.  Graphitic is defined as having the XRD lines of three-dimensional graphite.

\begin{figure}[htb!]
  \centering
\includegraphics[width=0.89\textwidth]{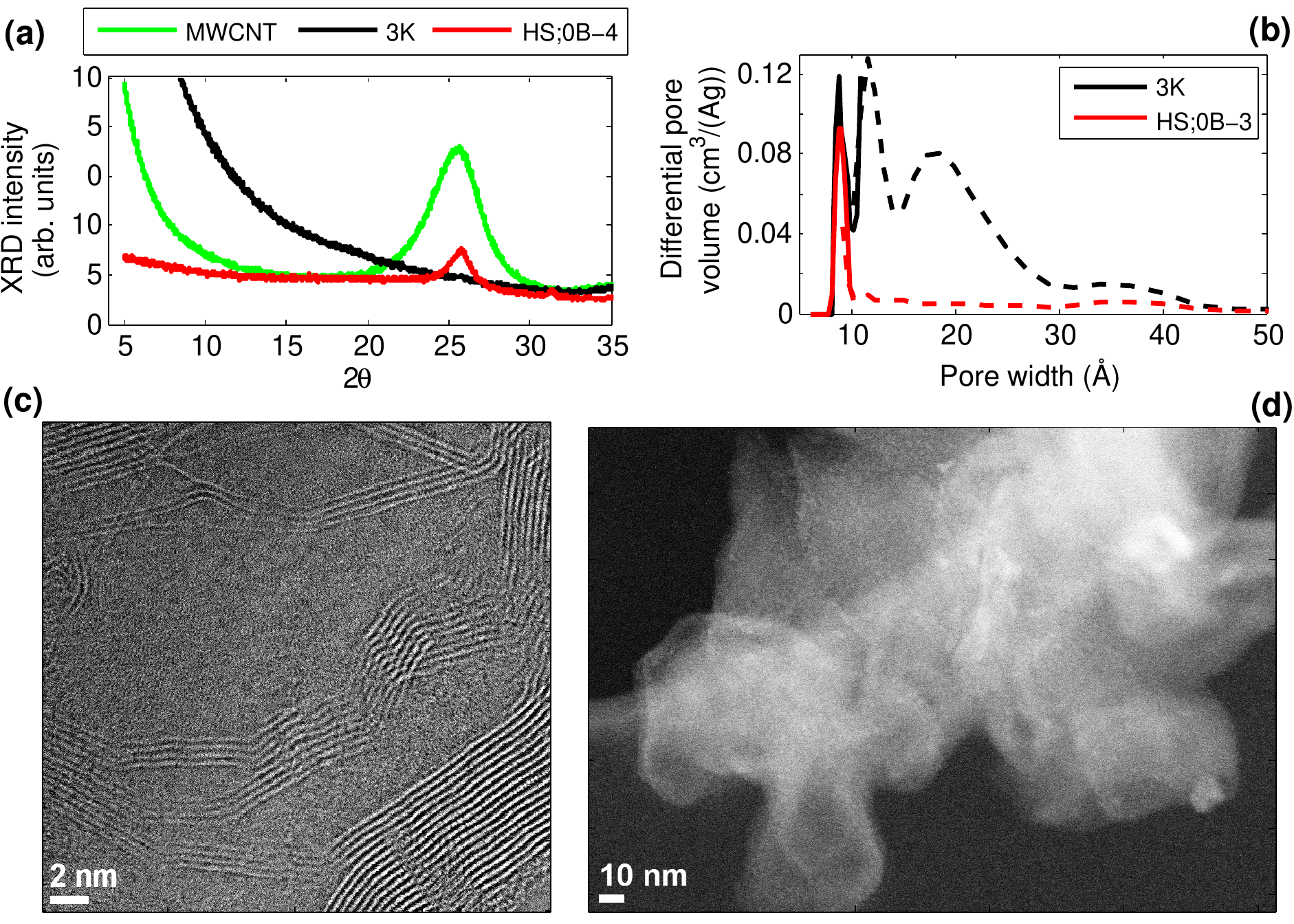}
\caption{Materials characteristics of carbon adsorbents.  (a) Powder X-ray diffraction spectra from sample of interest, `HS;0B-4' and control `3K', as well as a sample of multi-walled carbon nanotubes (MWCNTs).  The MWCNTs and `HS;0B-4' both show the 002 peak of graphite.  (b)  Pore size distributions calculated from CO$_2$ and N$_2$ adsorption isotherms. (c)  Bright field STEM image of `HS;0B-2', showing graphite in the material.  Large pores formed where graphene layers split apart can be seen.  (d)  Dark field STEM image of `HS;0B-2', showing a flakey grain of the material lying on the sample grid.    }
\label{mat}
\end{figure}

Published pore size distributions of  'HS;0B-\#' variants \cite{IINScarb,thesis} (where \# refers to different batches synthesized by the same method\cite{thesis}) have shown that the samples have a narrow distribution of pores under 1 nm in width while `3K' has a broad range of pore sizes with an average size over 1 nm.  Further characterization data are shown in the Extended Data (ED) Fig. \ref{mat}.    X-ray diffraction spectra  were collected from powdered samples using Cu K-$\alpha$ X-rays.  Microscopy images were collected using an aberration corrected JEOL 2200FS 200 keV scanning transmission electron microscope.  Pore size distributions were calculated from CO$_2$ isotherms and N$_2$ isotherms measured with a Quantachrome Autosorb-1 instrument.\cite{umc}  The TEM image in Fig \ref{mat}(c) show regions where graphene stacks into nanoscale regions of graphite, resulting in the broad 002 graphite peak observed in the XRD spectrum in Fig. \ref{mat}(a).  But the other XRD lines of three-dimensional graphite are not observed, thus we refer to the `HS;0B' as `locally graphitic,' though it is a non-graphitizable carbon.

We have previously shown inelastic neutron scattering data from H$_2$ in variants of the same samples \cite{IINScarb} at low incident neutron energies, 30 meV and 90 meV, and at lower temperatures and pressures, $T$=23 K and $P$\textless50 mbar.  The present measurements were conducted using similar methods  for sample outgassing to remove oxygen and water, loading of  carbon samples into cylindrical aluminum sample cans under $^4$He atmosphere with quartz wool between carbon powders and seals, loading of the sample can into the instrument and \emph{in situ} leak checking with helium, flushing of helium and addition of hydrogen, and data collection.   Here we describe in detail only methods that differ significantly, which primarily involve the different thermodynamic conditions.  

Neutron measurements were performed using the MARI spectrometer at the ISIS Neutron Source.  Except where noted, the incident neutron energy was 400 meV.  After outgassing and cooling,  background spectra from the can+carbon+wool were collected at $T=15$, 84, and 101 K.   The volumes in the system consisted of 1.7 cc (cm$^3$) in the sample can at the measurement $T$, 3.2 cc of capillary line with a gradient between $T$ and room temperature, and 72 cc at room temperature where pressure was continually monitored with an accuracy of $\pm 1$ bar.   After background measurements, the system was pressurized by connecting the system to a gas cylinder containing high purity room temperature normal-H$_2$ (an equilibrium mixture of para-H$_2$ and ortho-H$_2$) at $P$\textgreater200 bar with the sample cell at $T$=101 K, then the system was isolated from the gas cylinder and the sample cooled to 52 K.  Reductions of pressure were conducted by venting the system to atmosphere for $P>1.2$, and to a vacuum system for $P<1.2$.

As spectra were collected, the temperature was varied between $T$=52--101 K  with a constant amount of H$_2$ in the system.  We report here the pressure of these isochores at $T$=101 K as the nominal pressure.  The system was allowed to equilibrate for 1.5 hours after addition of H$_2$, and 0.5 hours after temperature changes before spectra were collected.  Data was collected for approximately 2 hours at each $T$, $P$.  The temporal order of the collected spectra was: `HS;0B-3' background,  `HS;0B-3' at nominal H$_2$  pressures of 187 bar then 30 bar then 123 bar, `3K' background,  then `3K' at 186 bar, always from low to high $T$ when $P$\textgreater0.  Given this temporal ordering, this means that we observed two transitions from a normal phase of  H$_2$ to the novel high mass recoil phase, and one transition in the other direction, with all these transitions observed to occur solely in response to changes in the temperature of the sample $T$.  Both transitions to the novel phase were observed within the same $T$ range, but at quite different $P$, and the second transition into the novel phase was observed after 29 hours in the normal phase.  After the second set of measurements in the novel phase, the sample was changed to the control `3K', in which only the normal phase was observed.

The pressure of the room temperature volume rose with temperature between every data point, and the final pressure at $T$=101 K after measurements was within $\pm$2 bar of the loading pressure, thus there is no evidence of a leak in the system.  This equilibriation with the room temperature volume also means that the amount of hydrogen in the sample can decreased slightly as its temperature increased.  Besides the oxygen ($M$=16.0 amu) in the quartz wool, no significant amount of any element with a mass near 18.1 amu, including fluorine ($M$=19.0 amu) or neon ($M$=20.2 amu), was known to be present during synthesis\cite{thesis} or preparation of the samples, or in the  instrument vicinity.   The sample mounting stick in the instrument was not adjusted during measurements, nor do we have any reason to suspect that normal operation of the instrument would result in significant motion of the sample into or out of the beam.  For each spectrum collected from the system containing hydrogen, the background spectrum collected at the closest temperature was used for background subtraction.  Analysis of high $Q$ aluminum Bragg peaks was used to estimate self-shielding factors $f$ used for background subtraction.  A minimum value of $f=0.81$ was found in the normal phase at $P$=187 bar and $T$=52 K, and a minimum of $f=0.66$ was found in the novel high mass recoil phase at $P$=187 bar and $T$=75 K. 

\clearpage

\section{Extended Data}

\begin{figure}[htb!]
  \centering
\includegraphics[width=0.49\textwidth]{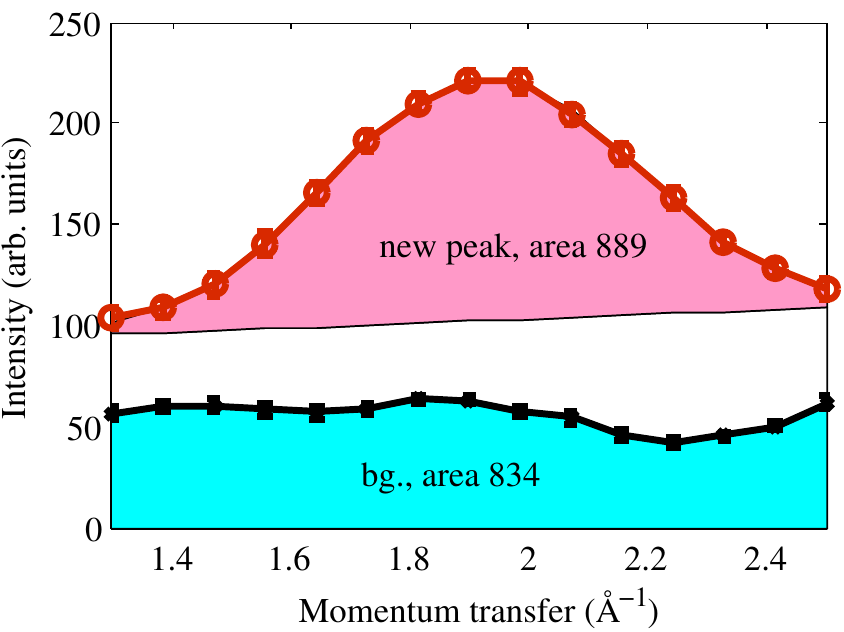}
\caption{Data from the elastic line for H$_2$ in `HS;0B-3' summed over $\hbar \omega=\pm 10$ meV at $T=61$ K, $P=187$ bar, with background subtraction, shown in comparison to the background.  Areas under the background and the novel elastic peak are shown, with the latter obtained by fitting a linear background plus a Guassian to the elastic scattering with the area under the Gaussian given.}
\label{fitel}
\end{figure}

 \begin{figure}[htb!]
  \centering
\includegraphics[width=0.49\textwidth]{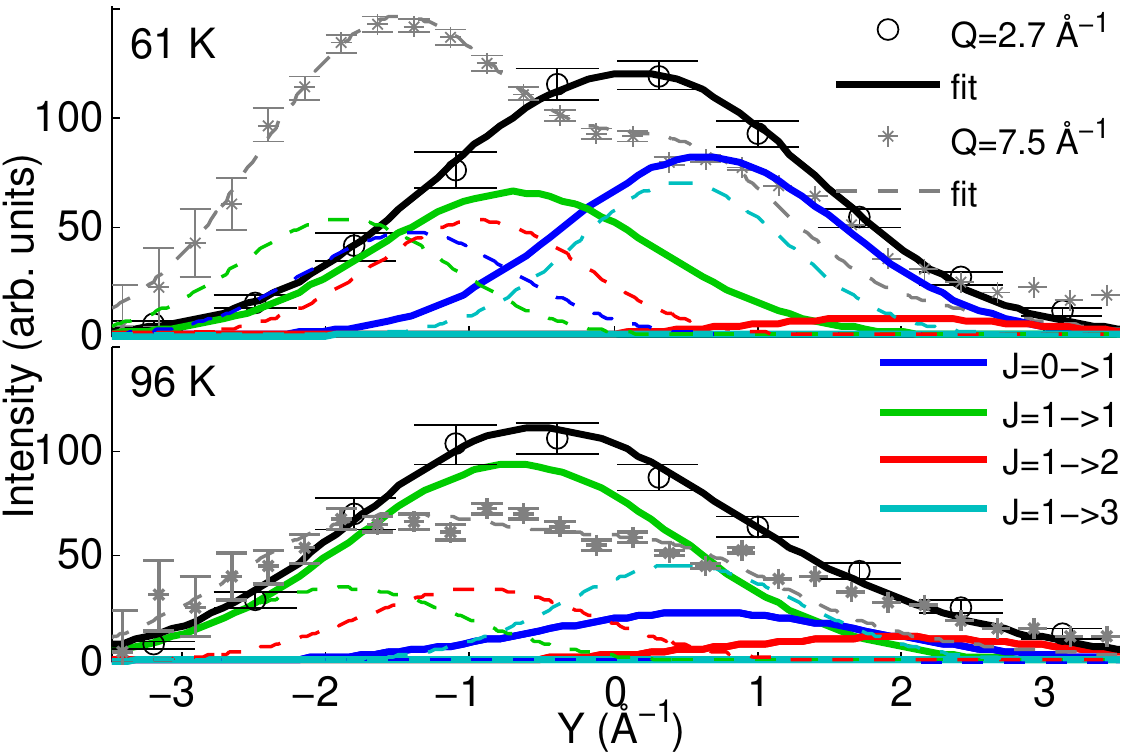}
\caption{Inelastic neutron scattering data reparametrized as $J^H(Y^H,Q)$ for several fixed values of $Q$.  Spectra are collected from H$_2$ adsorbed in `HS;0B-3' at a nominal pressure of 187 bar and temperatures of 61 K (top) and 96 K (bottom).  Data are fit using an H$_2$ molecular recoil model, which has separate components for each rotational transition.  The separate rotational components are shown using colored lines of the same linetype as the total fit (thick solid and thin dashed for $Q$=2.7, 8.3 \AA$^{-1}$ respectively).  }
\label{fitf}
\end{figure}

 \begin{figure}[htb!]
  \centering
\includegraphics[width=0.49\textwidth]{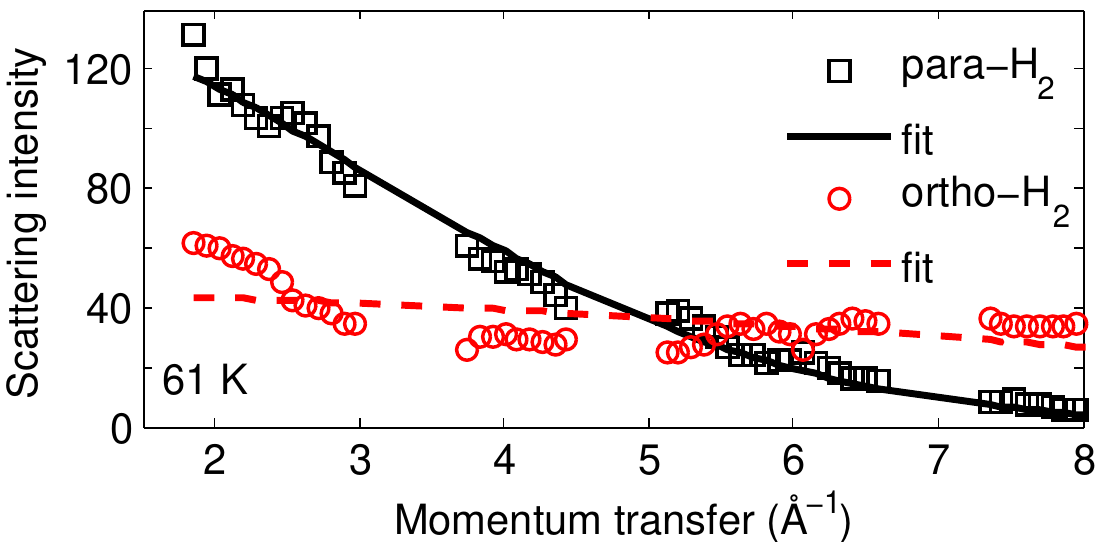}
\caption{Intensity of inelastic neutron scattering from ortho-H$_2$ and para-H$_2$ as a function of momentum transfer in the sample of interest at P=187 bar,  T=61 K.  The results indicate the Debye-Waller factor of the para-H$_2$ is much larger than that of the ortho-H$_2$, meaning the ortho-hydrogen has less thermal motion, consistent with it being more strongly confined and thus likely preferentially adsorbed in the pores
 }
\label{opfit}
\end{figure}

\begin{figure}[htb!]
  \centering
\includegraphics[width=0.49\textwidth]{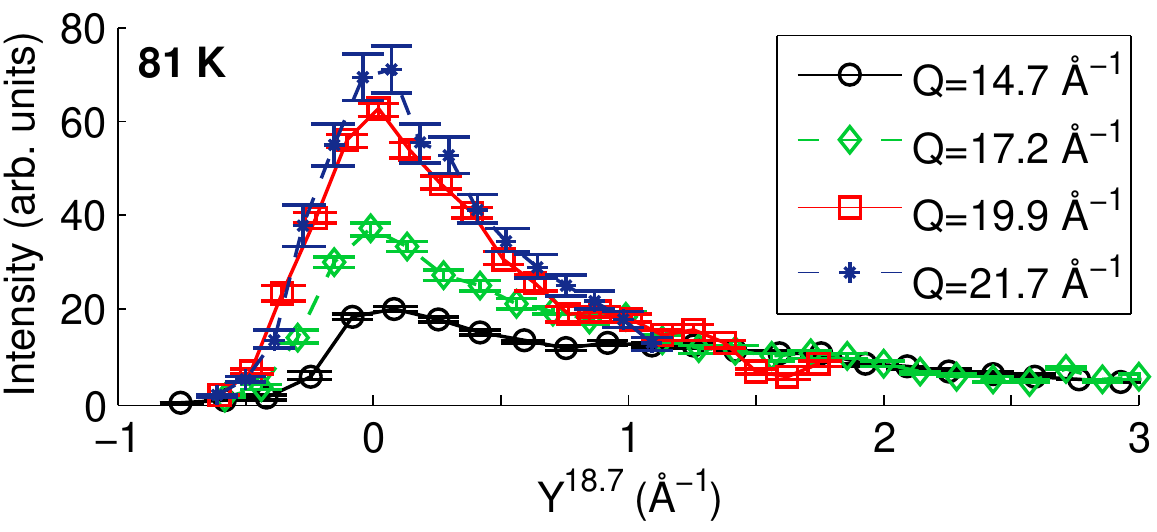}
\caption{Inelastic neutron scattering data reparametrized as $J(Q,Y^{18.1})$ for several fixed values of $Q$ adsorbed in `HS;0B-3' at a nominal pressure of 187 bar and temperature of 81 K.   }
\label{Y18}
\end{figure}

\clearpage

\section{Supplementary Discussion}

\subsection{Alternative explanations for measured INS spectra}

Here we will discuss in more detail possible explanations for the two observed spectral changes reported in the main text, which include an intense novel peak which appears on the elastic line and a change in the recoil mass from 1.2 amu to 18.1 amu.  These changes are observed at temperatures between 74--92 K only at high H$_2$ pressure, $P=123,187$ bar from H$_2$ adsorbed in the sample of interest, `HS;0B-3.'  These spectral features are not observed at any temperature at low pressure, $P=30$ bar, or from H$_2$ adsorbed in the control sample `3K'; under these conditions only normal scattering as expected from uncorrelated H$_2$ was observed.

In the main text, we briefly conclude that these changes originate from a change in phase of the H$_2$ in the system.  We further associate this novel phase with one in which nuclear spin coupling between ortho-H$_2$ ($S=1,J=1$)  creates the observed elastic peak and antiferromagnetic ordering of the 9 accessible internal sub-states of ortho-H$_2$ results in collective recoil of 9 H$_2$, which have a combined mass of 18.1 amu.

Here we will consider any possible alternative explanations for these observed spectral changes, which are summarized in Table \ref{alt}. 

\begin{table}[thb!]
\caption{Alternative explanations for the change in neutron scattering spectra observed at pressures $>$120 bar and temperatures between 74--92 K from H$_2$ adsorbed in the sample of interest `HS;0B-3', which is a graphitic carbon with narrow pores.}
  \centering
  \begin{tabular}{p{4.5cm}|p{6cm}|p{6cm}}
    \toprule
    \textbf{Observed spectral change} & \textbf{Alternative explanation} & \textbf{Contraindicated by} \\
    \midrule
    1.~Change in recoil mass from 1.2 to 18.1 amu &\tabitem Formation and recoil of any 18.1 amu molecule or complex, such as H$_2$O or methane & \tabitem As explained in main text, recoil mass is equal to atomic mass as long as nuclear spins are uncoupled  \\
& & \tabitem Chemical reaction unlikely to be reversible at observed temperatures  \\
&\tabitem ~18 amu atom added to/removed from system  & \tabitem Transition preserves total scattering intensity; incoherent neutron scattering cross section of $H$ is 8360 times larger than the incoherent cross section of any element with a mass close to 18.1 amu (O, F and Ne).      \\
&  & \tabitem Assymetric DINS scattering, shown in Fig. \ref{Y18} and discussed below in Section \ref{highm}, is not explained by an atomic model   \\
& & \tabitem The intensity of DINS scattering in the novel phase shown in Fig. \ref{Y18} also grows with $Q$ in defiance of the zeroth sum rule (Eq. \ref {sum0}), thus is also not explained by an atomic model \\
& \tabitem Multiple scattering, combination of high $|\vec{Q}_1|$ elastic scattering event and low $|\vec{Q}_2|$, $\hbar\omega$ recoil scattering event &  \tabitem Multiple scattering is usually diffuse and creates no additional peaks because $|\vec{Q}_1$+$\vec{Q}_2|$ may represent constructive or destructive addition, while observed 18.1 amy recoil scattering has a distinct peak $\hbar \omega$ at each $Q$ (see Fig. \ref{Y18}) \\
&& \tabitem No strong Bragg peak above $Q>18$ \AA$^{-1}$ (Fig. \ref{bg}), while high M recoil observed up to 37 \AA$^{-1}$ (Fig. \ref{H218}(a)). \\
&& \tabitem No explanation for observed temperature/sample dependence \\
    2.~Novel elastic peak appearing at 1.9 \AA$^{-1}$ & \tabitem High H$_2$ pressure squeezes graphene planes together, increases intensity of broad graphite 002 peak at 1.87 \AA$^{-1}$ in background & \tabitem Large intensity of novel elastic peak relative to background (Fig. \ref{fitel}) means that all of the carbon would need to be converted to graphite\\
&& \tabitem Saran carbons are non-graphitizable at temperatures up to 2000 $^\circ$C (see Methods), indicating that restructuring is a very high energy process which is unlikely to be reversible \\
&& \tabitem Graphitization of structures shown in Fig. \ref{mat} requires anisotropic pressure, whereas gas pressure is isotropic\\

& &\tabitem  No explanation for observed temperature dependence   \\
&\tabitem Material with intense Bragg peak moved into/out of beam& \tabitem No such material known to be in the vicinity \\
&& \tabitem No other new elastic peaks observed \\
&& \tabitem  No explanation for observed temperature/pressure dependence \\
    \bottomrule
  \end{tabular}
\label{alt}
\end{table}

To explain the experimental data, it is important that the physical process which results in the observed changes in neutron scattering be \emph{reversible}.  During the course of the experiment, we passed between the normal phase and the novel phase multiple times, collecting a total of 11 spectra in the novel phase and 20 spectra in the normal phase.  Furthermore, we collected data from H$_2$ in `HS;0B-3' first at 187 bar, observing both the normal and novel phase.  We then spent 29 hours collecting data at 30 bar, in which only the normal phase was observed.  We then increased the H$_2$ pressure to 123 bar, where the transition to the novel phase was observed again within the same temperature range.  We then changed the sample, after which only the normal phase was observed.  

This temporal ordering, the observation of the novel phase in many spectra with small errorbars, and the consistent temperature range at two pressures make these results unlikely to be caused by some fluke of instrumentation.  We are not aware of any issues with the MARI spectrometer which have ever before produced spectral changes such as these.  

None of the alternative explanations summarized in Table \ref{alt} are able to convincingly explain one of the observed spectral changes, let alone both.  Thus we conclude that the most likely explanation is nuclear spin correlations between the H$_2$ molecules, as described in the main text.

\subsection{Fits of the H recoil in the normal phase}

Uncorrelated H$_2$ molecules have been well-studied with neutron scattering.  Here we discuss fits to our neutron scattering data, showing that the measured scattering in the normal phase is consistent with that of uncorrelated H$_2$.  (Spectra in the novel phase are not fit, as no theoretical model for this phase exists.)  We also show evidence that ortho-H$_2$ is preferrentially adsorbed within the pores of the carbon material.

The neutron scattering method used in this work is known as deep inelastic neutron scattering (DINS), and is typically used to measure momentum distributions of nuclei.  `Deep' refers to the high energy of the incident neutrons, which allows one to penetrate deeply into the system under study.  The simplest interpretation of DINS spectra uses the impulse approximation (IA), which assumes that the neutron interacts instantaneously with a single nucleus, transferring energy and momentum and causing the nucleus to freely recoil away as if it were unaffected by the other particles in the system.

An atomic nucleus with mass $M$ and initial momentum $\vec{p}$ has a momentum of $\vec{p}+\vec{Q}$ after interacting with the incident neutron.  Making the assumption of the IA that the change in energy of the nucleus can be approximated by the change in kinetic energy because the scattering happens over such short time scales that the change in potential energy is small, the energy transfer is $\hbar \omega = \hbar \omega_R^M + \vec{p}\cdot \hbar \vec{Q}/M$, where $\hbar \omega_R^M=(\hbar Q)^2/2M$ is the recoil energy.   $S_I(\vec{Q},\omega)$ is then given by 
\begin{equation}
S_I(\vec{Q},\omega)=\int n(\vec{p_i}) \delta \left(\hbar \omega - \hbar \omega_R^M - \frac{ \vec{p}\cdot \hbar \vec{Q}}{M}\right) d^3 \vec{p}.
\end{equation}
where $n(\vec{p})$ is the distribution of momenta in the system.  At a given $\vec{Q}$, the result is a peak centered at the recoil energy, $\hbar \omega_R^M$, and broadened by the distribution of initial momenta in the system.  Defining an IA scaling variable, 
\begin{equation}
 \label{y} Y^M(Q,\omega)= \frac{M}{\hbar^2 Q} \left( \hbar \omega - \hbar \omega_R^M  \right),
\end{equation}
one obtains,
\begin{eqnarray}
\label{comp} 
S_I(\vec{Q},Y^M)&=&\frac{M}{\hbar Q}\int n(\vec{p_i}) \delta (\hbar Y -  \vec{p}_i \cdot  \hat{Q}) d^3 \vec{p_i}. \\
\nonumber &=& \frac{M}{\hbar Q} J(Y^M,\hat{Q}),
\end{eqnarray}
where  $\hat{Q}$ is a unit vector.  Thus as $Q$ increases, one expects the maximum scattering intensity to fall off as $Q^{-1}$, and the full width at half maximum (FWHM) to increase linearly with $Q$.

The IA is only exact at infinite energy and momentum transfer, or practically when the incident neutron energy is larger than the system binding energies  and $2 \pi/ Q$ is smaller than the distances between nuclei.  At lower values of $Q$, the details of DINS spectra depend on the local environment of the nuclei.  When the nuclei are part of a fluid of molecules, DINS spectra can be better described as free recoil of the molecular center of mass plus excitations of the internal rotational degrees of freedom.  One must also take into account the correlations between the proton spins of the two H atoms in the molecule.   We use this well-known model of DINS scattering from uncorrelated H$_2$ molecules with mass $M^{H_2}$=2.0 amu \cite{YandK,greatFit,langel1988inelastic,bafile1998deep} to fit our data in the normal phase using, 
\begin{eqnarray}
\label{yH2} 
Y_{JJ'}^{H_2}(\hbar \omega)&=& \frac{M^{H_2}}{\hbar^2 Q} \left( \hbar \omega - \hbar \omega_R^{M^{H_2}}- (E_J'-E_J)  \right)\\
\label{fit} S(\hbar \omega)&=& P \sum_{J'} F_{0J}^{H_2}(Q) e^{-(Y_{1J'}^{H_2}(\hbar \omega))^2/2 \sigma^2} \\
 \nonumber &+&O \sum_{J'} F_{1J}^{H_2}(Q)  e^{-(Y_{0J'}^{H_2}(\hbar \omega))^2/2 \sigma^2}  ,
\end{eqnarray}
where $Y_{JJ'}^{H_2}$ is the IA scaling variable for a hydrogen molecule ($M^{H_{2}}$=2.0 amu) undergoing a a transition from the molecular rotational state $J$ to $J'$, and $F_{JJ'}^{H_2}(Q)$ is the scattering form factor for the $J{\rightarrow} J'$ transition \cite{YandK} which includes the appropriate coherent $\sigma_C$ and incoherent $\sigma_I$ scattering cross sections depending on the initial and final spin of the transition.  The only free parameters in this fit are $P$, which is proportional to the amount of para-hydrogen with intial spin $S=0$, $O$, which is proportional to the amount of ortho-hydrogen with initial spin $S=1$, and $\sigma$, which gives the spread of the momentum distribution.  This model assumes that due to the low temperature, which is smaller than the spacing between the rotational states, only the $J$=0,1 rotational states are initially occupied.

Results of the fit are shown for several spectra at two values of $Q$ in Fig. \ref{fitf}.  In Figure \ref{opfit} we show $P$ and $O$ as a function of $Q$ for H$_2$ adsorbed in the sample of interest `HS;0B-3' at a pressure of 187 bar and temperature of 61 K, which is just below the observed transition to the novel phase.  The scattering intensity $P$ is proportional to $N_P$, the number of para-H$_2$ molecules in the system, but is attenuated by the thermal motion of the molecules through the Debye-Waller factor, 
\begin{eqnarray}
P(Q)=N_P e^{-Q^2 D_P}, \\
O(Q)=N_O e^{-Q^2 D_O}.
\end{eqnarray}
From Fig. \ref{opfit}, one can clearly see that the Debye-Waller factor of the two species is different, with $D_O<D_P$.  Since the Debye-Waller factor is proportional to the amount of thermal motion of the atoms, this result seems to indicate that the ortho-H$_2$ is more strongly confined such that it undergoes less thermal motion.  This is consistent with the ortho-H$_2$ being preferrentially adsorbed in the pores of the carbon material, and the para-H$_2$ more likely to be in the bulk gas phase which exists between the micrometer sized grains of the carbon.

\subsection{Properties of the high M recoil in the novel phase}
\label{highm}

 In Figure \ref{Y18} we show $J(Q,Y^{18.1})$ for several values of $Q$, where $J$ and $Y^M$ are  are defined as in Eqs. \ref{y} and \ref{comp} with $M$=18.1.  Under this scaling, data which represents single-particle self-correlations should be symmetrically broadened peaks centered at $Y$=0 which become nearly identical as $Q$ increases.    In contrast, Fig. \ref{Y18} shows that the intensity of $J(Q,Y^{18.1})$ grows strongly with $Q$; indeed a gross violation of the zeroth moment sum rule. The peak is centered at $Y$=0, meaning the recoil mass of 18.1 amu describes the peak location, but the peak is very asymmetric and the degree of asymmetry does not decrease with $Q$.  

Rather, it is curious to note that within the long asymmetric tail at $Y^{18.1}$\textgreater1, the spectra $J(Y^{18.1},Q)$ for different values of $Q$ do lie on top of one another as predicted by the IA, while  the intensity grows with $Q$ only for $|Y^{18.1}|\lesssim$1.  In any case, the observed scattering certainly does not fit the predictions of the IA for single atom recoil.

\end{document}